\documentclass[reprint,showpacs,preprintnumbers,amsmath,amssymb,prl]{revtex4-1}
\usepackage{amsmath}
\usepackage{amsfonts}
\usepackage{amsthm}
\usepackage{dsfont}
\usepackage{graphics}
\usepackage{graphicx,bm}
\usepackage[caption=false]{subfig}
\usepackage{amssymb}
\usepackage{mathrsfs}
\usepackage{braket}
\usepackage{float}
\usepackage{color}
\usepackage{url}
\usepackage[abs]{overpic}
\usepackage[usenames,dvipsnames]{xcolor}
\usepackage{commath}
\usepackage{subfig}
\usepackage{multirow}
\usepackage{verbatim}
\usepackage{array}
\usepackage[hidelinks]{hyperref}

\begin{document}

\title{Simulating Non Commutative Geometry with Quantum Walks}

\author{Fabrice Debbasch$^{1}$}
\email{fabrice.debbasch@gmail.com}

\affiliation{
%{$^{1}$Physics Department, The University of Western Australia, Perth, WA 6009, Australia}\\
{$^{1}$Sorbonne Universit\'e, Observatoire de Paris, Universit\'e PSL, CNRS, LERMA, F-75005, {\sl Paris}, France}}

\date{\today}
\begin{abstract}
Non Commutative Geometry (NCG) is considered in the context of a charged particle moving in a uniform magnetic field. The classical and quantum mechanical treatments are revisited and a new marker of NCG is introduced. This marker is then used to investigate NCG in magnetic Quantum Walks. It is proven that these walks exhibit NCG at and near the continuum limit.
For the purely discrete regime, two illustrative walks of different complexities are studied in full detail. The most complex walk does exhibit NCG but the simplest, most degenerate one does not. Thus, NCG can be simulated by QWs, not only in the continuum limit, but also in the purely discrete regime. 
%Non Commutative Geometry (NCG) arising in the Landau problem of an electron moving in a uniform and constant magnetic field is first revisited in the context of classical and quantum mechanics without spin. The treatment is then extended to magnetic Dirac Quantum Walks (DQWs), which can be realised with photonics. The walks are first studied at and near their continuum limit (Dirac equation), then in the purely discrete regime. Two illustrative examples are worked out in detail. They show that NCG always appear at and near the continuum limit, generally appears in the discrete regime, but may be absent from some degenerate cases.
%Thus, photonics can simulate NCG through QWs.

\end{abstract}
%\pacs{03.67.-a, 47.37.+q, 47.40.-x, 67.10.-j}
\maketitle

%\section*{Popular Summary}
%
%Quantum simulation was first suggested by Feynman and aims at reproducing key behaviour of complex quantum systems through other, less complex but also quantum systems which can be realised and observed in a laboratory. Quantum Walks (QWs) are discrete automata. They are quantum analogues of classical random walks and have been realised in the laboratory. They are important in quantum information because they are a universal computational primitive {\sl i.e} any quantum 
%algorithm can be expressed as a QW.  This article demonstrates that quantum walks can be used to simulate Non Commutative Geometry (NCG), which arises in physical contexts as diverse as the quantum Hall effect and quantum gravity.
%
%The simulation of NCG with QWs rests on two key observations. First, the simplest example of NCG in physics can be found is  in the dynamics of a particle coupled to a magnetic field (Landau problem). Second, this dynamics can be simulated with quantum walks. We show in the article that usual, continuous NCG can thus be simulated by considering QWs at their continuum limit. We also show that QWs, considered in their discrete regime away from their continuum limit generate at new, purely discrete and richer NCG.  
%
%The next step of this work is to analyse systematically the continuous and discrete NCGs generated by QWs and to also explore their possible use in quantum algorithmics.

\section{I. Introduction}

Quantum Walks (QWs) are automata defined on graphs and lattices. They were first considered by Feynman in studying possible discretizations for the Dirac path integral \cite{feynman2010quantum,schweber1986feynman}. They were later introduced in a systematic way by Aharonov et al. \cite{aharonov1993quantum} and Myers \cite{meyer1996quantum}.
%and are simple quantum automata defined on 
%lattices and graphs. 
QWs are useful in quantum information and algorithmic development \cite{ambainis2007quantum,magniez2011search,ManouchehriWang2014} because they are a universal computational primitive.%and are powerful tool in
%Quantum walks are simple models of coherent quantum transport on discrete structures such as graphs and lattices. 
They are also important for quantum simulation \cite{Getal10a} and they have been realised experimentally in a number of ways \cite {ManouchehriWang2014} which include cold atoms \cite{Ketal09a}, photonic systems \cite{Petal10a,Setal10a} and trapped ions \cite{HAetal20a}. 

 There is now a growing literature on the geometrical aspects of QWs. QWs can indeed be used to simulate Dirac fermions interacting with arbitrary Yang-Mills gauge fields \cite{arnault2016quantum,marquez2018electromagnetic} and arbitrary relativistic gravitational fields \cite{di2013quantum, di2014quantum, AFF16a,AF17a,arnault2017quantum}. In particular, exact discrete gauge invariant Yang-Mills field strengths can be built with QWs \cite{arnault2016quantum2}, and one can even construct a discrete counterpart to the Riemann curvature tensor \cite{D19b}. Also, discussions of the Lorentz invariance of QWs can be found in \cite{AFF14a,BDAP17a,D19a,ABDAP19a} and the symplectic discrete geometry behind  some QWs has been presented in \cite{D19a}.

Now, what about Non Commutative Geometry (NCG)? NCG, which has been introduced more than seventy years 
ago \cite{S47a}, has become an important part of modern physics, arising in contexts as diverse as the quantum Hall effect \cite{BESB94a}, fluid dynamics \cite{BC06a,DG16a}, string theory and M-theory \cite{SW99a}. Yet, there is no discussion in the literature of a possible quantum simulation of NCG. 

This is all the more surprising because the simplest example of NCG in physics can be found in the so-called 
Landau problem \cite{J02a, M03a, AM04a}, which describes the dynamics of a point particle (electron) in a uniform and constant magnetic field, and QWs have been proposed to simulate this problem \cite{arnault2016landau}, together with possible experimental realisations through cold atoms \cite{Sajidetal19a}. 

The aim of the present article is to show that QWs can be used to perform a quantum simulation of NCG. 
We focus on the Landau problem and start by revisiting NCG, first in the context of classical mechanics, then in the context of
non relativistic spinless quantum mechanics. Addressing these relatively simple situations allows us to develop an intuition about NCG in the Landau problem and we also set up the tools that will be useful in the rest of the article. We then switch to QWs, 
focusing on thee so-called magnetic QWs, which can simulate the relativistic quantum Landau problem.  
%The material is organised as follows. Though QWs are essentially discrete models of spin $1/2$ quantum relativistic dynamics, we start our presentation by revisiting the spinless Landau problem in both classical and non relativistic quantum mechanics. 
%
%
%The appearance of NCG is these contexts is much easier the understand tha
%
%This presents the advantage, not only of making the article completely self-contained, but also 
%
%for spinless non relativistic particles
%%in the contexts of continuous classical and quantum mechanics, 
%and develop an approach which can be generalised to deal with QWs. We then switch to magnetic QWs, 
%These walks are studied, first at and near their continuum limit, then in the purely discrete regime. 
We show that magnetic QWs at and near the continuum limit
%, which coincides with the Dirac equation, 
do realise NCG. We then explore the purely discrete regime through two relatively simple examples where all computations can be carried out at least semi-numerically, if not analytically. The most complex walk does exhibit NCG but the simplest, most degenerate one does not. Moreover, the discrete NCG, when it occurs, displays a non-linearity and a purely local (as opposed to global) character that the continuous NCG does not. We finally summarise and discuss all results in the last section. The general conclusion is that QWs, both in their continuum and purely discrete regimes, can simulate NCG.

\section{II. Spinless particle}

\subsection{II.1. Classical mechanics}

Consider the planar motion of a point charge $e$ of mass $m$ submitted to a constant and uniform magnetic field $\bf B$ perpendicular to the plane of motion. The equations of motion read
\begin{eqnarray}
{\ddot x} & = & \omega {\dot y} \nonumber \\
{\ddot y} & =  - & \omega {\dot x} 
\end{eqnarray}
where $x$ and $y$ are two orthonormal coordinates in the plane, $\omega = e B/m$ $m$ and a derivation with respect to the time $t$ is designated by a dot. 
The complex velocity $V = {\dot x} + i {\dot y}$ obeys ${\dot V} = - i \omega V$, which leads to
\begin{equation}
V(t) = V_0 \exp(- i \omega t),
\label{eq:Vt}
\end{equation}
where $V_0 = v_{0x} + i v_{0y}$ is the initial complex velocity. Integrating the expression for $V(t)$ delivers
\begin{equation}
X(t) = - i \frac{V_0}{\omega} \exp(- i \omega t) + X_c,
\label{eq:Xt}
\end{equation}
where $X(t) = x(t) + i y(t)$ and $X_c = x_c + i y_c$ is a complex integration constant. Equation (\ref{eq:Xt}) leads immediately to 
\begin{equation}
\mid X(t) - X_c \mid^2 = \left( x(t) - x_c  \right)^2 + \left( y(t) - y_c  \right)^2 = R^2
\end{equation}
with $R = \mid V_0 \mid/\omega$. The variable $X(t)$ oscillates harmonically around its time-average $X_c$ and 
the trajectory of the particle is a circle of radius $R$ centred on $(x_c, y_c)$.

It is possible to consider that $X(t)$ is 
approximately equal to $X_c$ at all times if $\mid X(t) - X_c\mid/\mid X_c \mid \ll 1$ {\sl i.e.}
if the amplitude of the oscillations is very small compared to $X_c$.. This translates into $\mid V_0 \mid /\omega \ll  \mid X_c \mid$.

The Lagrangian $L$ and the Hamiltonian $H$ for a particle moving in an electromagnetic field involve the vector potential $A$ \cite{LL75a}, so writing $L$ and $H$ makes it necessary to choose a gauge for $A$. We choose the so-called longitudinal gauge, where $A_x = 0$ and $A_y = - B x$, and the Lagrangian then reads
\begin{equation}
L(x, y, {\dot x}, {\dot y}) = \frac{m}{2} \, \left(
{\dot x}^2 + {\dot y}^2
\right)
+ e B x {\dot y}.
\end{equation}
The momenta conjugate to $x$ and $y$ are $p_x = m {\dot x}$ and $p_y = m {\dot y} + e B x$, and the Hamiltonian reads
\begin{equation}
H(x, y, p_x, p_y) = \frac{1}{2m} \left(
p_x^2 + (p_y - e B x)^2
\right).
\end{equation}

As shown previously, the $2$-degrees of freedom Hamiltonian system is integrable. It thus admits at least two independent first integrals. All trajectories are actually bounded and periodic (as opposed to pseudo-periodic), so there are actually three independent first integrals. The first of these is obvious, it is the time-independent Hamiltonian, which represents the conserved kinetic energy of the particle. The second integral is nearly as obvious, but its interpretation is not. Since $L$ (and $H$) do not depend explicitly on $y$, the momentum $p_y$ is conserved. Using the integrated expressions for $V$ and $X$ given above shows that, on the motion, $p_y$ coincides with $ eB x_c = m \omega x_c$. The final first integral is $p_x + m \omega y$, which coincides on the motion with $m \omega y_c$. 

%Non commutative geometry {\sl i.e.} non commutation of $x$ and $y$ appears in two different guises. The first one involves averaging. 

%Let now $f$ and $g$ be two differentiable functions defined on the $4$d phase-space. Their Poisson bracket is defined in the usual manner by
%\begin{eqnarray}
%\{f, g\} & = & \frac{\partial f}{\partial p_x}\, \frac{\partial g}{\partial x} + \frac{\partial f}{\partial p_y}\, \frac{\partial g}{\partial y} \nonumber \\
%& - & \frac{\partial f}{\partial x}\, \frac{\partial g}{\partial p_x} - \frac{\partial f}{\partial y}\, \frac{\partial g}{\partial p_y}.
%\end{eqnarray}
%This definition entails ${p_x, x} = {p_y, y} = 1$, so $p_x$ and $x$ on one hand and $p_y$ and $y$ are non commuting variables in the classical sense.
Non commutative geometry {\sl i.e.} non commutation of $x$ and $y$ appears when one approximates the circular trajectory of the particle by its center
{\sl i.e.} neglects the oscillations in the variable $X$. Consider indeed the Lagrangian ${\tilde L}$ defined by
\begin{equation}
{\tilde L}(x, y, {\dot x}, {\dot y}) = \frac{m}{2} \,
{\dot x}^2 + e B x {\dot y},
\end{equation}
which can be obtained from $L$ by neglecting $\frac{m}{2} \,{\dot y}^2$ compared to $\frac{m}{2} \,{\dot x}^2$. The momentum ${\tilde p}_x$ is identical to 
$p_x = m {\dot x}$ but ${\tilde p}_y = e B x = m \omega x$. This relation is a first class constraint. It makes the $x$ variable proportional to ${\tilde p_y}$, which 
has a non vanishing Poisson bracket with $y$. If one were to follow Dirac's procedure to quantize the dynamics of $\tilde L$, the first class constraint 
would have to be enforced on the physically admissible states, ensuring that, for these states, the operator $x$ is always proportional to the operator
${\tilde p}_y$, which does not commute with the operator $y$. 
%
%I MISS SOMETHING ABOUT THE LOSS OF ONE DIMENSION AND THE CHOICE THAT IS MADE OF KEEPING THE NON COMMUTATION INSTEAD OF THE COMMUTATION. WHY???
%
%THE ANSWER IS: FIRST CLASS CONSTRAINT!
%
%
%
%
%which leads to $\{x, y\} = \{{\tilde p}_y, y\}/(m \omega) = 1(m \omega)$. 
The equations of motion derived from $\tilde L$ reveal how this non commutative geometry relates to neglecting the spatial oscillations of the variable $X(t)$. The $y$-equation of motion derived from $\tilde L$ is $\dot {\tilde p}_y = 0$, which implies $\dot x = 0$, so $x = x_c$ and the $x$-equation of motion is $\dot {\tilde p}_x = e B {\dot y}$, which is equivalent to $m {\ddot x} = eB {\dot y}$. Since $x = x_c$, this implies ${\dot y} = 0$ {\sl ie.} $y = y_c$.

Let us end this discussion by pointing out that the point $x_C + i y_C$ is actually the exact time average of $X(t)$. We will see in the next section that NCG in the quantum mechanical treatment of the Landau problem can also appear in two different, but related manners, one which relies on neglecting the width of a wave-function, which is the quantum mechanical equivalent of neglecting the classical amplitude of oscillations, and one which relies on an averaging, and which does not entail any approximation.

%$ e B x {\dot y}$. On the motion, this comes down to neglecting
%$\frac{m}{2} \, v_{0y}^2$ compared to $eB v_{0x} v_{0y}/ \omega$ {\sl i.e.}

%Introducing $L$, $H$ and the momenta $p_x$ and $p_y$ illuminates the geometry of the problem 

%The general solution of these equations reads
%\begin{eqnarray}
%x(t) & = & \frac{1}{\omega}\ 
%\left( 
%v_{0x} \sin(\omega t) - v_{0y} \cos(\omega t)
%\right) + x_c \nonumber \\
%y(t) & = & \frac{1}{\omega}\ 
%\left( 
%- v_{0x} \cos(\omega t) - v_{0y} \sin(\omega t)
%\right) + y_c
%\end{eqnarray}
%where  $v_{0x}$ and $v_{0y}$ are initial velocity components and $x_c$ and $y_c$ are two other integration constants. The trajectory of the particle is a

\subsection{2.2 Quantum mechanics}

Coordinates, momentum components and Hamiltonian become the operators $\hat x$, $\hat y$, ${\hat p}_x$, ${\hat p}_y$ and $\hat H$. 
The Hamiltonian ${\hat H}$ commutes with ${\hat p}_y$ because
${\hat p}_y$ generates translations in the $y$ direction and ${\hat H}$ is independent of ${\hat y}$. 
We therefore search for eigenstates of both ${\hat H}$ and
${\hat p}_y$ and denote them by $\mid E, p >$.  As well-known, the spectrum of ${\hat p}_y$ is the set of real numbers $\mathbb R$. For each real number $p$, let ${\mathcal H}_p$ be the subspace spanned by vectors of the $\mid ..., p>$. On ${\mathcal H}_p$, the Hamiltonian operator coincides with ${\hat H}_p$ given by
\begin{equation}
{\hat H}_p = \frac{1}{2m} 
{\hat p}_x^2 +  \frac{m \omega^2}{2} 
({\hat x} - {\hat x}(p))^2
\end{equation}
where $\omega = eB/m$ (as in the previous section) and ${\hat x}(p)$ is the operator which, in the $x$-representation, coincides with the multiplication by $x(p) = p/(eB) = p/(m \omega)$.

The reduced Hamiltonian ${\hat H}_p$ thus generates the dynamics of an harmonic oscillator of mass $m$ and pulsation $\omega$, centred on $x(p)$. The energy eigenstates 
of ${\hat H}_p$ are non-degenerate, indexed by $n \in {\mathbb N}$, with eigenvalues $(E_p)_n = (n + 1/2) \hbar \omega$ independent of $p$. The wave-function of the state
$\mid (E_p)_n, p >$ is the product of a Gaussian by a Hermite polynomial, both centred on $x(p)$. It is either symmetric (even $n$) or antisymmetric (uneven $n$) 
with respect to $x(p)$ and has typical width $\mid x - x(p) \mid \sim
 \left ((n + 1/2) \hbar/(m \omega) \right)^{1/2} = (n + 1/2)^{1/2} a$ where the length $a = \left(\hbar/(m \omega) \right)^{1/2}$ is also independent of $p$. The eigenvalue
 $p$ of $p_y$ thus enters the eigenstate of ${\hat H}_p$ only through the shift $x(p)$.
 
 In any given sate, the coordinate 
 %operator ${\hat x}$ 
 $x$
 can be viewed a random variable with probability law given by Born's law. 
% So is the commutator $\left[ {\hat y}, {\hat x} \right]$. 
 Since the eigenfunctions are either symmetric or antisymmetric with respect to $x(p)$, the average value of 
 % ${\hat x}$
 $x$
  in any eigenstate $\mid (E_p)_n, p >$ is identical to $x(p)$, which is proportional to the $y$-momentum $p$.
  Now, the average of $x$ is, be definition, the expectation value of the operator ${\hat x}$, which we denote 
  by $<{\hat  x} >$. Thus, averaging or taking the expectation value delivers non commutative geometry in the sense that, on the energy eigenstates, $ \left[ {\hat y}, <{\hat  x} > \right] = i \hbar/(m \omega)$, as can be checked directly on the wave-functions. This is true for all values of the parameters.
 
 Non commutative geometry also appears without averaging, but at the cost of an approximation. Suppose the width $\mid x - x(p) \mid$ of the wave-function is much smaller than $x(p)$ {\sl i.e.} if
 $\mid e B \mid \ll p^2/(\hbar (n + 1/2)^2)$. The probability law for $x$ is then essentially a Dirac distribution on $x(p)$, and 
 one gets $ \left[ {\hat y},  {\hat x} \right] \simeq i \hbar/(m \omega)$. 
 %Note that there is no condition on $eB$ for $<{\hat x}>$ not to commute with ${\hat y}$.

%\section{III. Dirac particle} 

\section{III. Quantum Walks}

\subsection{III.1 Magnetic Quantum Walks}

A discrete time QW simulating on a cartesian space-time grid the $2D$ motion of a particle immersed in a constant and uniform magnetic field has been proposed in \cite{arnault2016landau}.
Let $j \in \mathbb N$ be the discrete time and $(q, r) \in {\mathbb Z}^2$ be the discrete position on the $2D$ grid perpendicular to the magnetic field $B$. The equations defining the walk read:
\begin{eqnarray}
\psi^L_{j+1, q, r} & = & e^{2i \alpha_q} c^- \left[ c^+ \psi^L_{j, q+1, r+1}  + i s^+ \psi^R_{j, q-1, r+1}
\right] \nonumber \\
& & + i e^{-2i \alpha_q} s^- \left[ i s^+ \psi^L_{j, q+1, r-1}  +  c^+ \psi^R_{j, q-1, r-1}
\right] \nonumber \\
\psi^R_{j+1, q, r} & = & i e^{2i \alpha_q} s^- \left[ c^+ \psi^L_{j, q+1, r+1}  + i s^+ \psi^R_{j, q-1, r+1}
\right] \nonumber \\
& & + e^{-2i \alpha_q} c^- \left[ i s^+ \psi^L_{j, q+1, r-1}  +  c^+ \psi^R_{j, q-1, r-1}
\right] 
\end{eqnarray}
where $\psi^{L/R}$ are the two components of the `spinor', $\alpha_q = \epsilon^2 e B q/(2 \hbar)$, $c^\pm = \cos \theta^\pm$, $s^\pm = \sin \theta^\pm$ with $\theta^{\pm} = \pm \pi/4 - \epsilon m/(2 \hbar)$ and $\epsilon$ is a real positive parameter of the walk. An interesting continuum limit can be obtained by defining $t_j = \epsilon j$, $x_q = \epsilon q$, 
$y_r = \epsilon r$, assuming analyticity in $\epsilon$ and letting $\epsilon$ tend to $0$. This delivers
\begin{eqnarray}
\hbar(\partial_t  - \partial_x)\psi^L -i (\hbar \partial_y  + i e B x - m) \psi^R & = & 0 \nonumber\\
i (\hbar \partial_y  + i e B x + m) \psi^L +  \hbar (\partial_t  + \partial_x)\psi^R & = & 0,
\end{eqnarray}
which is the Dirac equation describing the planar motion of a spin $1/2$ charge $e$ immersed in the magnetic field $B$ orthogonal to the plane of motion. The velocity of light $c$ is set to unity in the whole article.  
 
To investigate if NCG appears, we search for stationary states of the QW (or of the Dirac dynamics) which are also eigenstates of 
the momentum operator ${\hat p}_y$. We thus write $\Psi^p_{j, q, r} = \Phi_{j, q} \exp(-i p \epsilon^2 r/\hbar)$ and obtain for $\Phi$ the equations:
\begin{eqnarray}
\phi^L_{j+1, q} & = & e^{2i \beta_q} c^- \left[ c^+ \phi^L_{j, q+1}  + i s^+ \phi^R_{j, q-1}
\right] \nonumber \\
& & + i e^{-2i \beta_q} s^- \left[ i s^+ \phi^L_{j, q+1}  +  c^+ \phi^R_{j, q-1}
\right] \nonumber \\
\phi^R_{j+1, q} & = & i e^{2i \beta_q} s^- \left[ c^+ \phi^L_{j, q+1}  + i s^+ \phi^R_{j, q-1}
\right] \nonumber \\
& & +  e^{-2i \beta_q} c^- \left[ i s^+ \phi^L_{j, q+1}  +  c^+ \phi^R_{j, q-1}
\right] 
\end{eqnarray}
with $\beta_q = \alpha_q -  \epsilon^2 p/(2 \hbar) = \epsilon eB (x_q - x(p))/(2 \hbar)$. The evolution of $\Phi$ is thus  a $p$-dependent QW on the line. Nevertheless, 
for readability purposes, the retained notation does not make the dependance of $\beta$ and $\Phi$ on $p$ explicit. At the continuum limit, the equations for $\Phi$ read:
\begin{eqnarray}
\hbar (\partial_t  - \partial_x)\phi^L + (e B (x - x(p)) + i m) \phi^R & = & 0 \nonumber\\
(- e B (x - x(p)) + i m) \phi^L + \hbar (\partial_t  + \partial_x)\phi^R & = & 0.
\end{eqnarray}
The expression of $\beta_q$ above, and the appearance of the difference $x - x(p)$ in the continuum limit Dirac equation
suggest that NCG is also present in the discrete time QW and, in particular, in the Dirac equation. To prove or disprove that intuition, one needs to look at the energy-eigenstates of the dynamics. This can be done exactly at the continuum limit and at first order around this limit, but there does not seem to be a general expression for these states, valid for all values of the walk parameters. We therefore start by a general presentation of NCG at the continuum limit and at first perturbation order around this limit, and then switch to examples which illustrate possible behaviours of the QW in the purely discrete regime.

\subsection{III.2 NCG at and near the continuous limit}

The 
%continuum limit 
energy eigenstates for $\Phi$ have been computed in \cite{arnault2016quantum} at first order 
in the perturbation parameter $\epsilon$ around the continuum limit $\epsilon = 0$. After a change of orthonormal basis in
spin space, the components of these eigenstates are, at zeroth and first order in $\epsilon$, even or uneven functions of $x - x(p)$, so the average $< x >$ is identical to $x(p)$. Thus, as for the spinless non relativistic quantum mechanical problem, $\left[{\hat y}, < {\hat x} >\right] = i \hbar/(m \omega)$ at zeroth and first order in $\epsilon$ {\sl i.e.} at and near the continuum limit.

%It has also been shown in \cite{arnault2016quantum} that, at first order in the perturbation parameter $\epsilon$, the components of the energy eigenstates 
%remain even or uneven functions of $x - x(p)$. 
%Thus, 
%what has been said above about the continuum limit remains valid at first order around this limit. In particular, 
%at first order around the continuum limit, one still has 
%So,
%$\left[{\hat y}, < {\hat x} >\right] = i \hbar/(m \omega)$ remains valid at first order in $\epsilon$. 

In the new spin basis, the typical spatial extensions of the components of the eigenstate labeled by $n$ are
$\left((n + 1/2) a \right)^{1/2}$ and $\left((n - 1/2)a \right)^{1/2}$ where $a = \hbar \omega$. Both these extensions are much smaller than $\mid x(p)\mid$ if $\mid eB \mid \ll p^2/\left( \hbar ( n + 1/2)^2\right)$, which is the same condition as the one found 
in the quantum mechanical treatment. If this is realised, then  $ \left[ {\hat y},  {\hat x} \right] \simeq i \hbar/(m \omega)$ and one 
gets again NCG.

Thus, at and near the continuum limit, QWs essentially behave as solutions of the Schr\"odinger equation, as far as NCG is concerned. In other words, and perhaps unexpectedly, taking spin and special relativity into account does not bring anything new to the NCG discussion, even at first order in $\epsilon$. Things however change when one considers NCG in the 
purely discrete regime.

\subsection{III.3 Purely discrete case}

We now consider situations where $\epsilon = 1$ and $e B/\hbar = 2 \pi/N$, where $N$ is a positive integer. The Hilbert space is then made of $N$-periodic functions on the $q$-axis and is therefore of dimension $2N$. Assuming the square grid infinite in the %$y$- or 
$r$-direction, the momentum ${\bar p} = p/\hbar$ lies in an interval of length $2 \pi$, say 
$(0, 2 \pi)$.
%$(-\pi, \pi)$.
Suppose now one wants to investigate NCG in QWs using the same type of approximation as the one used in the quantum mechanical treatment. One would have to consider walks whose extension in the $q$-variable is much smaller than $\bar p$. 
But the extension of a walk cannot be smaller than the grid step, which is unity, and $\bar p$ is never much larger than unity, because it is in $(0, 2 \pi)$. Thus, the approximation used in the quantum mechanical treatment never applies to discrete autoata like QWs. But the other way to exhibit NCG, which is based on averaging, does work, as can be seen in Example 1 below.

%CHANGE THE FOLLOWING SENTENCE!
%
%We also suppose $p$ is a positive or negative integer, so a shift of $p$ leaves the $q$-axis invariant. With these assumptions,  $2 \beta_q = $

\subsubsection{III.3.1 Example 1}

Consider a walk with vanishing mass $m$, so $\theta_+ = - \theta_- = \pi/4$. The equations of motion simplify into
\begin{eqnarray}
\phi^L_{j+1, q} & = &  \cos(2 \beta_q) \phi^L_{j, q+1}  +  \sin(2 \beta_q) \phi^R_{j, q-1} \nonumber \\
\phi^R_{j+1, q} & = & - \sin(2 \beta_q) \phi^L_{j, q+1}  +   \cos(2 \beta_q) \phi^R_{j, q-1}. \end{eqnarray}
The coefficients in these equations are $2 \pi$-periodic in ${\bar p}$.
% = p/\hbar$.

Let us now specialise to the first non-trivial case {\sl i.e.} $N = 3$ and choose $\left\{0, 1, 2\right\}$ as spatial periodicity set. 
The time-independent eigenvectors of the dynamics satisfy
\begin{eqnarray}
\lambda \phi^L_{0} & = &  \cos(2 \beta_0) \phi^L_{1}  +  \sin(2 \beta_0) \phi^R_{2} \nonumber \\
\lambda \phi^R_{0} & = & - \sin(2 \beta_0) \phi^L_{1}  +   \cos(2 \beta_0) \phi^R_{2}  \nonumber \\
\lambda \phi^L_{1} & = &  \cos(2 \beta_1) \phi^L_{2}  +  \sin(2 \beta_1) \phi^R_{0} \nonumber \\
\lambda \phi^R_{1} & = & - \sin(2 \beta_1) \phi^L_{2}  +   \cos(2 \beta_1) \phi^R_{0}  \nonumber \\
\lambda \phi^L_{2} & = &  \cos(2 \beta_2) \phi^L_{0}  +  \sin(2 \beta_2) \phi^R_{1} \nonumber \\
\lambda \phi^R_{2} & = & - \sin(2 \beta_2) \phi^L_{0}  +   \cos(2 \beta_2) \phi^R_{1}.
\end{eqnarray}
A direct computation reveals that the characteristic equation obeyed by $\lambda$ is
\begin{equation}
\lambda^6 - \frac{3}{4} \lambda^4 - \frac{\cos(3 {\bar p})}{2} \lambda^3  - \frac{3}{4} \lambda^2 + 1 = 0.
\end{equation}
The spectrum is thus $2 \pi/3$-periodic in $\bar p$. On the period $(0, 2 \pi/3)$, the spectrum is also symmetric with respect to ${\bar p} = \pi/3$. The dynamics is unitary, so all eigenvalues lie on the unit circle. Since all coefficients in the characteristic equation are real, the solutions appear as couples of complex conjugates $(\exp(i \theta), \exp(- i \theta))$ with, say, $\theta \in (0, \pi)$. A numerical computation shows that the characteristic equation admits generically $6$ distinct solutions on the unit circle, except for ${\bar p} = 0, \pi/3, 2 \pi/3, 4\pi/3$, where the equation only admits $5$ distinct solutions. Finally, the eigenvalue/vector problem is invariant under the following two transformations:
\begin{eqnarray}
p & \rightarrow &  p + 2 \pi/3, \Phi_0 \rightarrow  \Phi_1, \Phi_1 \rightarrow  \Phi_2, \Phi_2 \rightarrow  \Phi_0 \nonumber\\
p & \rightarrow  & p + 4 \pi/3, \Phi_0 \rightarrow  \Phi_2, \Phi_1 \rightarrow  \Phi_0, \Phi_2 \rightarrow  \Phi_1.
\end{eqnarray}%WHAT FOLLOWS IS WRONG!
%
%${\bar p} = 0$ and $p = \pi/3$, in which cases the equation admits only $5$ solutions, so one of them has multiplicity two. These are the only cases when the equation admit a real solution, $+1$ for ${\bar p} = 0$ and $-1$ for ${\bar p} = \pi/3$. 

We now focus, for each value of ${\bar p}$, on the eigenvalue corresponding to the smallest positive value $\theta_s$ of the angle $\theta$, which is the discrete equivalent of the lowest positive energy. Figure 1 presents $\theta_S$ as a function of $\bar p$ on the period $(0, 2 \pi/3)$. Each of the corresponding normalised eigen-vectors generates a probability law on the set $\left\{ 0, 1, 2 \right\}$, and this law can be used to define and compute the average $< q >$ as a function of $\bar p$. This function $A$ is $2 \pi$-periodic, as are the coefficients of the equations obeyed by $\Phi$. It is plotted in Figure 2. 

NCG follows from the above results by the following reasoning. Let $f$ by an arbitrary, possibly $q$-dependent function of $r$. Its discrete Fourier transform 
\begin{equation}
{\hat f} ({\bar p}) = \frac{1}{\sqrt{2 \pi}}\, \sum_r f_r \exp(-i {\bar p} r)
\end{equation}
is defined for ${\hat p} \in (0, 2 \pi)$, and the inverse transform reads
\begin{equation}
f_r = \frac{1}{\sqrt{2 \pi}}\, \int_0^{2 \pi} {\hat f}({\bar p}) \exp(i {\bar p} r) d{\bar p}.
\end{equation}c
It follows from the second of these equations that
\begin{eqnarray}
r f_r & = & \frac{1}{\sqrt{2 \pi}}\, \int_0^{2 \pi} {\hat f}({\bar p}) r \exp(i {\bar p} r) d{\bar p}
 \nonumber \\
& = &  \frac{1}{\sqrt{2 \pi}}\, \int_0^{2 \pi} {\hat f}({\bar p}) \left(- i \partial_{\bar p} \exp(i {\bar p} r) \right) d{\bar p} \nonumber \\
& = & \frac{1}{\sqrt{2 \pi}}\, \int_0^{2 \pi} \left(  i \frac{d{\hat f}}{d{\bar p}} \right) \exp(i {\bar p} r)  d{\bar p}
\end{eqnarray}
where an integration by parts and the identity ${\hat f}(2 \pi) \exp( i 2 \pi r) =  {\hat f}(0) \exp( i 0 \times r)$ have been used to obtain the last equation. This shows that the operator $\hat r$ is identical to the operator $  i d/d{\bar p}$ and that therefore
$\left[{\hat r}, {\hat {\bar p}} \right] =  i$ or, if one prefers, $\left[{\hat r}, {\hat p} \right] =  i \hbar$. Using $<q > = A({\bar p})$ then delivers
\begin{equation}
\left[{\hat r}, < {\hat q} > \right] =  i 
%\frac{d A}{d{\bar p}}_{\mid{{\hat {\bar p}}}}
A' ({\hat {\bar p}}).
\label{eq:NonComDiscr}
\end{equation}
As seen in Figure 2, the function $A$ and thus its derivative can be inverted locally, but not globally. Thus, locally, 
(\ref{eq:NonComDiscr}) transcribes into an equation of the form
\begin{equation}
\left[{\hat r}, < {\hat q} > \right] =  i B(< {\hat q} >)
%\frac{d A}{d{\bar p}}_{\mid{{\hat {\bar p}}}}
\label{eq:NonComDiscrLoc}
\end{equation}
where $B$ is a non-linear function. 

\begin{figure}[t]
\includegraphics[width=.65\linewidth]{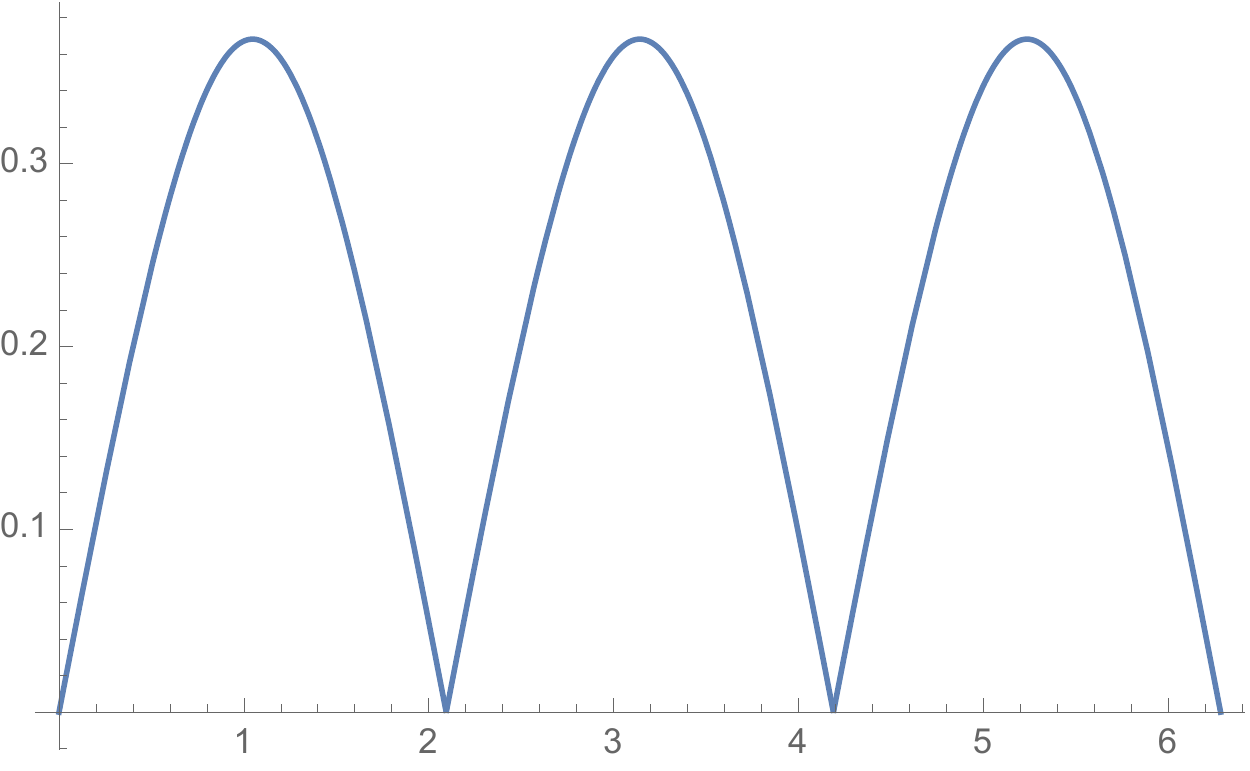}%
\caption{Example 1. Angle $\theta_S$ as a function of the reduced momentum variable $\bar p$.}
\label{ThetaS}
\end{figure}

\begin{figure}[t]
\includegraphics[width=.65\linewidth]{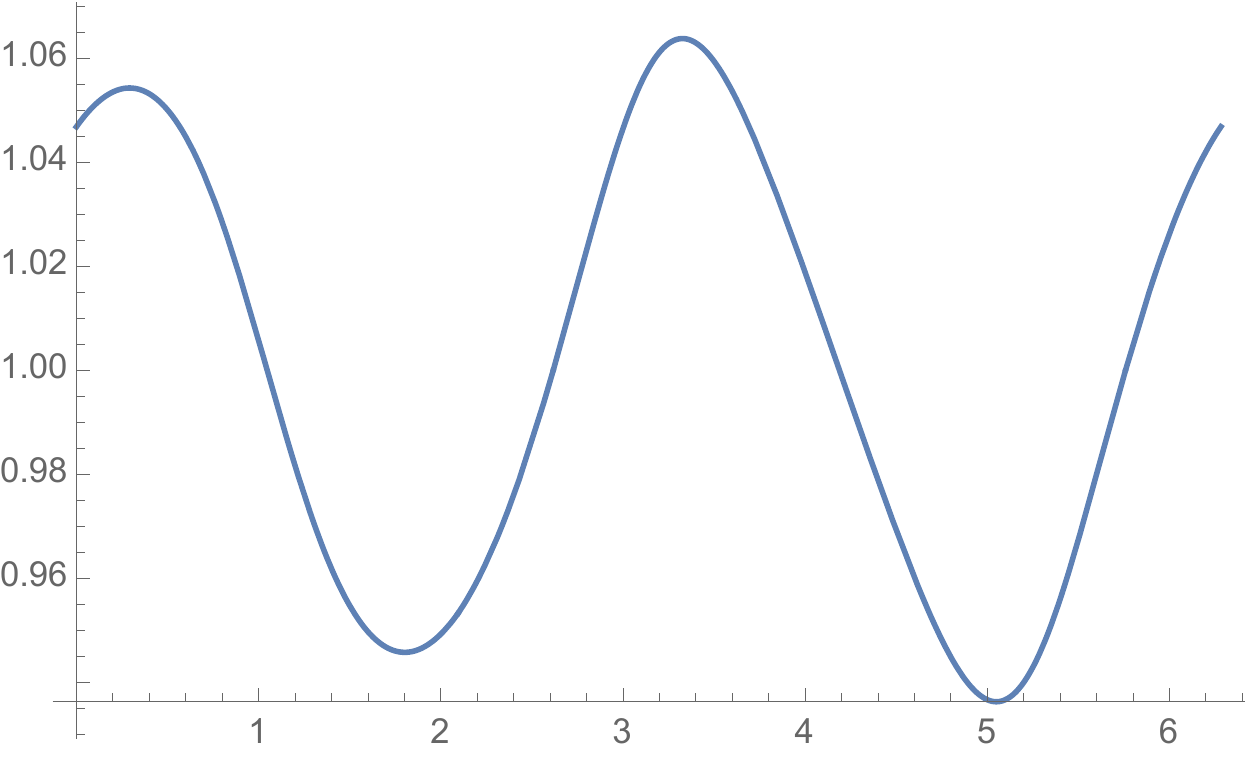}%
\caption{Example 1. Average position $< q >$ as a function of the reduced momentum variable $\bar p$.}
\label{Xav}
\end{figure}

%
%Table ... contains the value of $\theta_s$ for several values of ${\bar p}$.
%
%$0$: 0
%$\pi/24$: 0.0680494
%$\pi/12$: 0.134211
%$3\pi/24$: 0.196487
%$\pi/6$: 0.25268
%$5\pi/24$: 0.300366
%$\pi/4$: 0.33699
%$7\pi/24$: 0.360158
%$\pi/3$: 0.3681
%

\subsubsection{III.3.2 Example 2}

Consider now a walk with 
%$\epsilon = 1$ and 
$m/\hbar = \pi/2$, so that $\theta_+ = \pi/2$ and $\theta_- = 0$. The equations of motion for $\phi$ read
\begin{eqnarray}
\phi^L_{j+1, q} & = & i e^{2i \beta_q}  \phi^R_{j, q-1}
 \nonumber \\
  \phi^R_{j+1, q} & = & i e^{-2i \beta_q} \phi^L_{j, q+1}.\end{eqnarray}
 Choosing as spatial periodicity set $\left\{0, 1,  N-1\right\}$, the time-independent eigenvectors of the dynamics satisfy
 \begin{eqnarray}
 \lambda \phi^L_{0} & = & i e^{2i \beta_0}  \phi^R_{N-1}
 \nonumber \\
 \lambda  \phi^R_{N-1} & = & i e^{-2i \beta_{N-1}} \phi^L_0
 \nonumber \\
  \lambda \phi^L_{1} & = & i e^{2i \beta_1}  \phi^R_{0}
 \nonumber \\
 \lambda \phi^R_{0} & = & i e^{-2i \beta_0} \phi^L_{1}
 \nonumber \\
 \cdots \nonumber \\
  \lambda \phi^L_{N-1} & = & i e^{2i \beta_{N-1}}  \phi^R_{N-2}
 \nonumber \\
  \lambda \phi^R_{N-2} & = & i e^{-2i \beta_{N-2}} \phi^L_{N-1}
 \end{eqnarray}
 where $\lambda$ is the associated eigenvalue. 
 This system of $2N$ equations splits into $N$ independent systems of $2$ equations and each system of $2$ equations
 admits the same eigenvalues $\lambda_\pm = \pm e^{i \pi/N}$.
 % where $\Delta \beta = \beta_q - \beta_{q-1} = e B/ (2 \hbar)$ for all values of $q$. 
There are therefore only two energy levels $E_\pm = \hbar \pi (1/N + 1/2 \pm 1/2)$, which are independent of $p$, but each level is $N$ times degenerate. Each of the eigenvectors is localised on a couple of points $(q-1, q)$ with the convention that 
 $q - 1 = N - 1$ for $q = 0$. 
 The only non-vanishing components of the normalised eigenvectors localised at $(q-1, q)$ are
 \begin{eqnarray}
 \phi^L_q & = & \frac{1}{\sqrt{2}} \nonumber \\
 \phi^R_{q - 1} & = & \pm \frac{i}{\sqrt{2}} e^{i p/\hbar} e^{- i \pi \left( 1 + 2 (q -1)\right)/N},
 \end{eqnarray}
 with $q - 1 = N - 1$ if $q = 0$.
 %For example, $E_+$ admits as eigenvectors
 The eigenvectors do depend on the momentum $p$, but the associated probability densities do not. Thus, $p$ has no influence of the probability law for the position of the walker and non-commutative geometry in the sense of the previous sections is not realised for these walks.

%The energy eigenstates obey
%\begin{eqnarray}
%e^{i E /\hbar} \phi^L_{j, q} & = & i e^{2i \beta_q}  \phi^R_{j, q-1}
% \nonumber \\
%e^{i E /\hbar} \phi^R_{j, q} & = & - e^{-2i \beta_q} \phi^L_{j, q+1}.\end{eqnarray}
%The first of these equations implies
%\begin{equation}
%\phi^R_{j, q} = - i e^{- 2i \beta_{q+1}} e^{i E /\hbar} \phi^L_{j, q+1}
%\end{equation}
%which can be combined with the second equation to deliver
%\begin{equation}
% i e^{2 i E /\hbar} \phi^L_{j, q+1} =  e^{2i \pi/N} \phi^L_{j, q+1}.
%\end{equation}
%One finds similarly that 
%\begin{equation}
% i e^{2 i E /\hbar} \phi^R_{j, q-1} =  e^{2i \pi/N} \phi^R_{j, q-1}.
%\end{equation}
%Since $\phi^L$ and $\phi^R$ cannot both vanish, it follows form these equations that there are only two 
%eigen-energies, which obey
%\begin{equation}
%e^{ i E_\pm /\hbar} = \pm e^{i \pi(1/N - 1/4)} .
%\end{equation}

\section{IV. Discussion}

We have shown that NCG can be simulated with QWs. We have focused on the Landau problem, which describes the dynamics of a point particle in a uniform magnetic field. We have first revisited this problem in the contexts of classical mechanics 
and of spinless non relativistic quantum mechanics. We have then switched to so-called magnetic QWs, which model the relativistic dynamics of a spin $1/2$ quantum particle immersed in a magnetic field. 
%revisited NCG in the classical and quantum non relativistic Landau problem describing a spinless particle submitted to a uniform and constant magnetic field. We have introduced a new way to detect NCG in the classical and quantum mechanical problems, and we have used it on the so-called magnetic QWs, which can be realised through photonics. 
We have shown that 
NCG is always realised in magnetic DQWs at and near the continuum limit.  NCG can also be realised in the purely discrete regime, if the structure of the walk is rich enough. To show that NCG does not arise for all walks in the discrete regime, we have finally studied a specific, highly degenerate  magnetic walk where NCG does not occur. 
%in the discrete regime. 
The general conclusion is that NCG can be simulated with QWs, for example through quantum optics experiments. Let us now discuss all these results.

The treatment of NCG in the classical and quantum mechanical Landau problem proposed in this article differs from the existing ones in several respects. It has been known for a long time that NCG can arise through an approximation in the classical and quantum mechanical Landau problem, but previous presentations do not
% seem to
exhibit a dimensionless parameter controlling this approximation. We have identified this parameter and given its physical interpretation. As a consequence, we have shown that, at fixed value of the integer $n$ which indexes the oscillatory Landau levels, NCG arises for small enough values of the magnetic field, and not large values, as has been sometimes previously asserted based on heuristic reasoning. Note that the only previous non heuristic derivation of NCG in the classical and quantum mechanical Landau problem is based on restricting the Hilbert space allowed to the particle and this procedure does
not involve a dimensionless infinitesimal {\sl i.e.} it is a formal procedure which cannot be interpreted in usual physical 
terms as the consequence of neglecting a certain quantity with respect to another one having the same physical dimension.

We have also shown that
%% introduced 
%%a new (classical and quantum) observable, 
% the average position of the particle 
% as a new marker of
%% and shown that, in terms of this new observable, 
 NCG arises in the classical and quantum mechanical Landau problem independently of any approximation if one 
 considers the average position of the particle. This comes from the fact that, in all classical solutions and in the all quantum mechanical energy eigenstates, the average of one coordinate, say $x$, is proportional to the momentum associated to the other coordinate $y$. This new result is important for two reasons. First, because it shows that NCG is intrinsic to the Landau problem, and is not an approximate property of the dynamics valid only for small (or large) enough values of the magnetic field. Second, because it is this approach in terms of the average position operator that can be extended to encompass discrete automata like QWs. 

The extension to QWs at or near the continuum limit deserves only a few comments. The limit itself and its first order perturbations have already been studied in depth in \cite{arnault2016quantum}. The continuum limit obeys a Dirac equation. This equation can be reduced to the dynamics of two coupled harmonic oscillators and the components of the energy eigenstates are therefore 
very similar to the energy eigenstates of the quantum mechanical Landau problem. The discussion of NCG geometry therefore proceeds in a similar way. Considering first order perturbations around that continuum limit does not change drastically thee picture. In particular, the average position operator of one coordinate is still proportional to the momentum of the other coordinate.

Things however do change a lot when one considers magnetic QWs away from their continuum limit. The general analytical determination of the energy eigenstates does not seem feasible. We have therefore concentrated on two examples which correspond to values of the mass for which the equations of the walk simplify. The first example corresponds to a vanishing mass and is clearly the less degenerate of the two. In that case, NCG exists in the discrete regime, but it is much more complex than the NCG envisaged in the previous sections. Indeed, the average of one coordinate, say $x$ is again a function of the momentum $p$ associated to the other coordinate $y$, but this function is neither linear nor globally invertible. The commutator of the expectation value $<\hat x>$ and $\hat y$ is therefore not constant, but a function of $\hat p$, and this function can be inverted and written in terms of $<\hat x>$, but only locally. The NCG that emerges is thus quite different from the simplest one encountered in the continuous case. This new NCG is both discrete and non linear. It is controlled by the variable $\hat p$, which can be expressed in terms of $<\hat x>$ only locally. 

The second example is engineered to make the equations of motion of the walk as degenerate as possible. Because of the degeneracy, the wave-functions of the energy eigenstates depend exponentially on $p$, so the associated densities actually do not depend on $p$. It follows that the expectation value $<\hat x>$ does not depend on $\hat p$ and there is no NCG is this degenerate case.

Let us now conclude by mentioning a few possible extensions to this work. One should first concentrate on the same family of magnetic walks and determine which walks generate NCG. The examples worked out in this article suggest that NCG will appear in all walks except the most degenerate ones, but confirming this statement and investigating exactly what the NCG looks like for various values of the parameters, {\sl i.e.} for example how the non vanishing commutator depend on the impulse variable, will  clearly prove instructing. One should then consider other magnetic and electromagnetic walks, be they defined on the same grid, but with other unitaries, or be they defined on other grids, even possibly on graphs. Exploring how NCG appears in QWs simulating other physics than electromagnetism should also prove worthwhile. The results presented in this article that NCG also exists in other discrete models such as LGTs. If a discrete NCG does arise in LGTs and how it differs from the discrete NCG of QWs remains to be determined. Finally, the occurence of NCG in quantum walks, which are a universal computational primitive, makes one wonder about the consequences of NCG in quantum algorithmics.

\end{document}